# A weighted least squares procedure to approximate least absolute deviation estimation in time series with specific reference to infinite variance unit root problems

J. Martin van Zyl

Department of Mathematical Statistics and Actuarial Science, University of the Free State, PO Box 339, Bloemfontein, South Africa

**Abstract:** A weighted regression procedure is proposed for regression type problems where the innovations are heavy-tailed. This method approximates the least absolute regression method in large samples, and the main advantage will be if the sample is large and for problems with many independent variables. In such problems bootstrap methods must often be utilized to test hypotheses and especially in such a case this procedure has an advantage over least absolute regression. The procedure will be illustrated on first-order autoregressive problems, including the random walk. A bootstrap procedure is used to test the unit root hypothesis and good results were found.

1. **Introduction.**

An easy to calculate weighted least squares procedure is proposed which performs almost similar to least squares (LS) when the error term is Gaussian white noise. This estimation procedure performs very similar to least absolute deviation regression (LAD) in large samples with respect to bias and efficiency for series where the noise is heavy-tailed and from a stable distribution. In problems with large samples and many variables and especially where bootstrap procedures is used to test hypotheses the computational aspects of LAD regression can become a problem because of the computational complexity, and this method will have an advantage. The second advantage over LAD regression is that even in problems with small samples this procedure outperforms LAD regression in many cases with respect to bias and mean square error of the estimated parameters.

The procedure will be illustrated and tested on AR(1) models, with random walk models included. The principles involved will be similar in large autoregressive type regression problems where the errors are heavy tailed distributed and from a stable distribution.

Autoregressive models (AR) and the special case of a random walk play an important role in finance, for example in the efficient market hypothesis. In practice the distribution of the errors of the AR model is often heavy-tailed distributed with a finite mean and the variance may possibly not be finite, and not white noise as in the usual random walk or AR models. The case where the error term has a symmetric stable distribution with location parameter zero will be considered and specifically AR(1) models.

Hypotheses can be tested by making use of bootstrap methods based on the results of Moreno and Romo (2012), Davis and Wu (1997). The important result of Davis and Wu (1997), made it possible to test hypothesis using bootstrap samples, where a random normalizing constant for the test statistic in the case of heavy-tailed data was given, which can be used in the original sample and in bootstrap samples. It is shown in a simulation study that the unit root test can be tested with power comparable and better than much more computational complicated methods.

Results on identification and other properties of time series with heavy-tailed distributed errors were given in the paper of Adler, Feldman and Callagher (1998), Feigin and Resnick (1999). Subsampling Bootstrap, LAD estimation and hypothesis testing methods are considered in the papers of Moreno and Romo (2000), (2012), Davis and Wu (1997), Jach and Kokoska (2004). There is a huge literature about regular variation in time series. A few references of interest are the papers of Davies and Mikosch (1998). Li, Liang and Wu (2010), Andrews, Calder and Davis (2009), Chan and Zhang (2009), Dielman (2005), Samarakoon and Knight (2009).

Let $\{y_t\}, t = 1,....,T+1$ denote a time series and consider the general autoregressive model of order one (AR(1)):

$$y_t = \beta_0 + \beta_1 y_{t-1} + u_t, \qquad (1)$$

It will be assumed that the $u_t$'s has a stable distribution with characteristic function $\phi(t)$ where

$$\log \phi(t) = -\sigma^\alpha |t|^\alpha \{1 - i\beta sign(t) \tan(\pi\alpha/2)\} + i\mu t, \quad \alpha \neq 1,$$

and $\quad \log \phi(t) = -\sigma^\alpha |t|^\alpha \{1 - i\beta sign(t) \log(|t|)\} + i\mu t, \quad \alpha = 1.$

The parameters are the index $\alpha \in (0,2]$, scale parameter $\sigma > 0$, coefficient of skewness $\beta \in [-1,1]$ and mode $\mu$. It will be assumed that the errors are symmetrically distributed with $\mu = 0, \beta = 0$. The stable distribution has power tails of the form:

$$\Pr(|Y| > y) \propto y^{-\alpha}, y \to \infty, 0 < \alpha < 2,$$

with $E(y^r)$ finite for $r < \alpha$. Special cases of the stable distributions are for $\alpha = 1$, it is a Cauchy distribution and for $\alpha = 2$ a normal distribution.

Most of the theory of unit root models were derived assuming that the series of errors is white noise $\{u_t\}$, thus i.i.d. with mean zero and variance $\sigma_u^2$. In some financial applications, the $u_t$'s are called shocks or innovations. The parameter $\beta_0$ is often called the drift of the series. It can be shown that the unconditional mean and variance are $E(y_t) = \beta_0/(1-\beta_1)$ and $var(y_t) = \sigma_u^2/(1-\beta_1^2)$. For a random walk when $\beta_1 = 1$ and the mean and variance are both infinite, and the process is not stationary even if the series $\{u_t\}$ is white noise.

This hypothesis $H_0: \beta_1 = 1$ is called the unit root test of stationarity and the Dickey-Fuller test was developed to test the unit root test. Phillips (1987) derived results on the moment of such a process and showed for $\beta_0 = 0$, $\hat{\beta}_1$ is super consistent or $\lim_{T \to \infty} T(\hat{\beta}_1 - 1)$ has a well-defined limit.

If the $u_t$'s are from a stable distribution with index $\alpha$, $1 < \alpha < 2$, the usual estimators can be unstable. Van Zyl and Schall (2009) proposed an easy to calculate weighted estimation method and showed that for example this method outperforms the optimal trimmed mean proposed by (Rothenberg, Fisher and Tilanus (1964), Fama and Roll (1968)). A two stage estimation method is proposed, first plain least squares estimation is performed to estimate the residuals, calculate the weights using the residuals and then performing weighted least squares estimation. The weights proposed by Van Zyl and Schall (2009) are proportional to $\exp(-|\hat{u}_j - \hat{u}_m|)$, $\hat{u}_t, t = 1,...,T$, $\hat{u}_m$ is the sample median of the residuals.

AR(1) models including unit root models will be considered. With $\beta_1 = 1$ or $|\beta_1| < 1$, the noise, $\{u_t\}$ Gaussian white noise ($\alpha = 2.0$) or heavy-tailed with $1 < \alpha \leq 2$ a symmetrical stable distribution. A random walk where the increments are from a stable distribution with $\alpha > 1$, a finite mean and infinite variance is called a Lévy flight. In Lévy flights there are fewer large changes, but when there are changes, the changes are much larger than those which would be expected in the case with normally distributed shocks. Mandelbrot called this the Noah Joseph effect (Mandelbrot and Wallace (1968)).

## 2. Proposed estimation procedure

Robust estimation, making use of weights as in Van Zyl, Schall (2009) will be investigated. The weights are based on the idea of constructing a weighted mean which will be asymptotically equal to the median. In the papers of Koenker (2000), Portnoy (1997) for example, the problem where Laplace found LAD estimators using weighted medians is discussed. At the moment there is not yet a procedure where LAD problems can be solved in the form of a weighted sum. The proposed argument used is to approximate an integral which is equal to the median if the sample is from a symmetric distribution.

Consider the Laplace distribution with median $u_m$ and the integral

$$I = c\int_{-\infty}^{\infty} u e^{-|u-u_m|} du = c\int_{-\infty}^{\infty} (t+u_m)e^{-|t|} dt = u_m,$$

c a normalizing constant. Numerically the integral can be approximated by the sum $I \approx c\sum_{j=1}^{n} x_j e^{-|x_j - \hat{x}_m|}$, for a sample $x_1,...,x_n$ with estimated median $\hat{x}_m$, c a normalizing constant for the weights. If the differences $x_1 - \hat{x}_m,...,x_n - \hat{x}_m$ have an approximate Laplace distribution it can be seen that the sum approximate the median and in the case of an exact Laplace distribution will be equal to the median asymptotically in the limit. In practice it was found that this weighted sum outperforms the median in many problems where the data is heavy-tailed when estimating the parameter of location. The idea is to apply these weights to regression and in this work specifically to time series data.

In figure 1 the median of Cauchy data was estimated for various sample sizes using the median of the sample and also using the weighted mean estimator. It can be seen that the estimators are very close and the weighted estimator outperforms the median with respect to mean absolute deviation in most of the cases. At each sample size n, 500 samples were drawn and the mean deviation at each sample size for both estimation methods is plotted in figure 1.

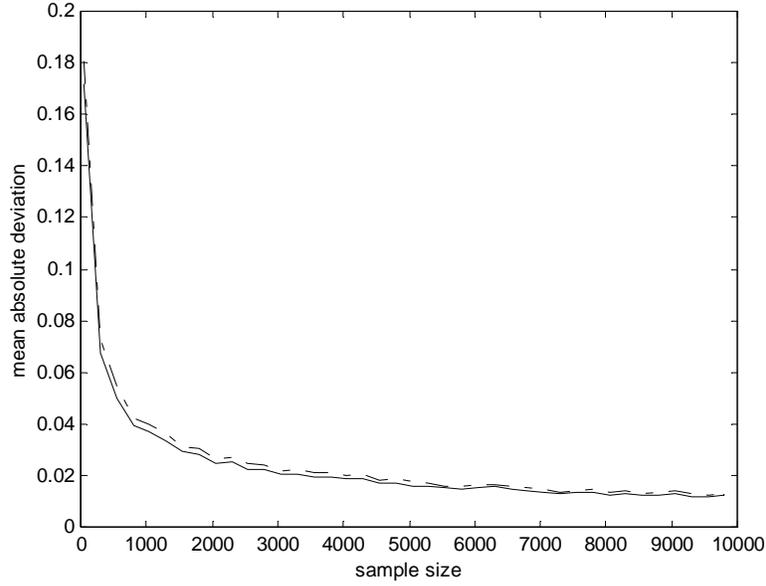

Figure1. Plot of mean absolute deviation of Cauchy distributed data using the median and the weighted mean method. The solid line denotes the average deviation for the weighted mean.

For an AR(p) model with constant of the form:

$$y_t = \beta_0 + \beta_1 y_{t-1} + ... + \beta_p y_{t-p} + u_t,$$

The errors $\{u_t\}, t = 1,...,T$, are symmetrically stable distributed with location parameter zero.

$$\mathbf{y} = \begin{pmatrix} y_{p+1} \\ y_{p+2} \\ \vdots \\ y_T \end{pmatrix}, \quad X = \begin{bmatrix} 1 & y_1 & \cdots & y_p \\ \multicolumn{4}{c}{................} \\ 1 & y_{T-p} & \cdots & y_T \end{bmatrix},$$

The usual estimate of the parameters using least squares is $\hat{\boldsymbol{\beta}} = (X'X)^{-1}X'\mathbf{y}$, and a series of estimated residuals can be calculated as $\hat{\mathbf{u}} = y - X\hat{\boldsymbol{\beta}}$. Calculate a series of weights

$$w_j = \exp(-|\hat{u}_j - \hat{u}_m|) / \sum_{j=1}^{T} \exp(-|\hat{u}_j - \hat{u}_m|), \quad w_j, \hat{u}_t, j = 1,...,T,$$

with $\hat{u}_m$ the sample median of the estimated residuals. Let W denote a matrix with diagonal the weights $w_1,...,w_T$, $w_j \geq 0, j=1,...,T, \sum_{j=1}^{T} w_j = 1$. The weighted estimate of the parameters is calculated as

$$\hat{\boldsymbol{\beta}}_w = (X'WX)^{-1} X'W\mathbf{y}.$$

For the AR(1) model the equations are

$$\begin{pmatrix} \hat{\beta}_{0w} \\ \hat{\beta}_{1w} \end{pmatrix} = \begin{pmatrix} 1 & \sum_{j=2}^{T} w_{j-1} y_{j-1} \\ \sum_{j=2}^{T} w_{j-1} y_{j-1} & \sum_{j=2}^{T} w_{j-1} y_{j-1}^2 \end{pmatrix}^{-1} \begin{pmatrix} \sum_{j=2}^{T} w_{j-1} y_j \\ \sum_{j=2}^{T} w_{j-1} y_{j-1} y_j \end{pmatrix}.$$

Experimentation showed that a few iterations improve the estimation results slightly in especially cases where the error terms are heavy-tailed.

### 3. Simulation study and application

#### 3.1 Estimation results

In the next simulation, the variances of the least squares and the weighted estimates were compared. AR(1) series, with T=250 points each were simulated n=1000 times. There is a drift with $\beta_0 = 0.1$. Let $\hat{\beta}_0, \hat{\beta}_1$ denote the usual least squares estimates and denote the weighted estimates by $\hat{\beta}_{0w}, \hat{\beta}_{1w}$ and LAD estimated parameters by $\hat{\beta}_{0L}, \hat{\beta}_{1L}$. The following table shows the estimated values, 1000 series of 250 each. The drift $\beta_0 = 0.1$ was used. The scale parameter $\sigma = 1.0$, symmetric stable errors with location parameter zero was simulated for the error term of the series. In the case where $\alpha = 2.0$ iterations did not improve results, otherwise two iterations were used in the weighted regression section. For $\alpha = 1.1$ there were cases with singular W matrices, and corrections can be made, for example as in ridge regression, keeping the constraint that the weights add up to one.

| $\beta_1$ | $\alpha$ | $\hat{\beta}_0$ | $\hat{\beta}_{0w}$ | $\hat{\beta}_{0L}$ | $\hat{\beta}_1$ | $\hat{\beta}_{1w}$ | $\hat{\beta}_{1L}$ |
|---|---|---|---|---|---|---|---|
| 1 | 2 | 0.1410 (0.1074) | 0.1428 (0.1232) | 0.1449 (0.1102) | -0.0152 (0.0005) | -0.0162 (0.0005) | -0.0162 (0.0006) |
| 1 | 1.9 | 0.1335 (0.3519) | 0.1106 (0.1384) | 0.1189 (0.1043) | -0.0165 (0.5258) | -0.0139 (0.4274) | -0.0141 (0.4159) |
| 1 | 1.5 | 0.2871 (5.3932) | 0.0883 (0.2527) | 0.0647 (0.1040) | -0.0192 (0.7113) | -0.0077 (0.1439) | -0.0074 (0.1459) |
| 1 | 1.3 | 0.3610 (102.2016) | -0.0148 (1.3982) | 0.0400 (0.2306) | -0.0189 (0.7111) | -0.0016 (0.0246) | 0.0038 (0.0505) |
| 1 | 1.1 | -1.4299 (8003.3) | -1.3741 (5684.9) | 0.0112 (0.6384) | -0.0203 (0.0014) | -0.0081 (0.0002) | -0.0019 (0.0000) |
| 0.975 | 2 | 0.0638 (0.0365) | 0.0605 (0.0457) | 0.0625 (0.0378) | -0.0180 (0.7546) | -0.0180 (0.9750) | -0.0180 (0.7819) |
| 0.975 | 1.9 | 0.0751 (0.0458) | 0.0600 (0.0417) | 0.0628 (0.0350) | -0.0187 (0.7746) | -0.0159 (0.8193) | -0.0166 (0.6879) |
| 0.975 | 1.5 | -0.0674 (13.6425) | -0.0003 (0.0206) | 0.0086 (0.0233) | -0.0174 (0.7262) | -0.0040 (0.1597) | -0.0070 (0.1891) |
| 0.975 | 1.3 | 0.0565 (5.0762) | 0.0068 (0.0183) | 0.0109 (0.0228) | -0.0164 (0.7091) | -0.0014 (0.0511) | -0.0039 (0.0769) |
| 0.975 | 1.1 | 1.2677 (448.0537) | 0.0099 (0.0465) | 0.0095 (0.0209) | -0.0153 (0.6404) | -0.0066 (0.0116) | -0.0017 (0.0235) |
| 0.5 | 2 | 0.0062 (0.0087) | 0.1081 (0.0143) | 0.0070 (0.0098) | -0.0129 (3.2210) | -0.0145 (5.8641) | -0.0136 (3.6482) |
| 0.5 | 1.9 | -0.0075 (0.0143) | -0.0033 (0.0129) | -0.0049 (0.0094) | -0.0105 (2.9646) | -0.0089 (3.9530) | -0.0049 (2.6555) |
| 0.5 | 1.5 | 0.0322 (0.2571) | 0.0013 (0.0126) | 0.0030 (0.0110) | -0.0104 (2.6331) | -0.0050 (1.1704) | -0.0053 (1.0433) |
| 0.5 | 1.3 | -.0216 (2.2063) | -0.0059 (0.0108) | -0.0033 (0.0113) | -0.0087 (2.2784) | -0.0015 (0.4960) | -0.0027 (0.4364) |
| 0.5 | 1.1 | -2.7439 (7455.4) | 0.0041 (0.0100) | 0.0057 (0.0123) | -0.0083 (1.9501) | -0.0005 (0.1513) | -0.0013 (0.1650) |

Table 1. Bias and MSE in brackets of the estimated parameters of an AR(1) and stable distributed innovations with different index values, using least squares and weighted least squares. Based on 1000 simulated series.

The estimation of $\beta_1$ using weighted LS compares favorably with the LAD estimation procedure. For $\alpha$ close to 1.00 and $\beta_1 = 1.00$, signs of multicollinearity was seen when using the weighted least squares. Ridge regression can help in those cases.

A comparison of the estimation of $\beta_1$ in the unit root model $y_t = 0.1 + \beta_1 y_{t-1} + u_t$, $\beta_1 = 1$ for different sample sizes, is shown in table 2. The errors are from a symmetric stable distribution with index $\alpha = 1.5, \sigma = 1.0$. The results are based on 1000 simulated series of length T each.

| T | $\hat{\beta}_{1w}$ | Bias($\hat{\beta}_{1w}$) | MSE($\hat{\beta}_{1w}$) | $\hat{\beta}_{1L}$ | Bias($\hat{\beta}_{1L}$) | MSE($\hat{\beta}_{1L}$) |
|---|---|---|---|---|---|---|
| 50 | 0.9586 | -0.0414 | 0.0058 | 0.9418 | -0.0582 | 0.0076 |
| 75 | 0.9771 | -0.0229 | 0.0026 | 0.9651 | -0.0348 | 0.0030 |
| 100 | 0.9851 | -0.0149 | 0.0009 | 0.9758 | -0.0242 | 0.0015 |
| 125 | 0.9889 | -0.0111 | 0.0005 | 0.9816 | -0.0184 | 0.0008 |
| 150 | 0.9912 | -0.0088 | 0.0003 | 0.9856 | -0.0144 | 0.0005 |
| 175 | 0.9930 | -0.0070 | 0.0002 | 0.9888 | -0.0112 | 0.0003 |
| 200 | 0.9948 | -0.0052 | 0.0002 | 0.9906 | -0.0094 | 0.0002 |
| 225 | 0.9946 | -0.0054 | 0.0001 | 0.9911 | -0.0089 | 0.0002 |
| 250 | 0.9960 | -0.0040 | 0.0001 | 0.9931 | -0.0069 | 0.0001 |
| 300 | 0.9969 | -0.0031 | 0.0001 | 0.9947 | -0.0053 | 0.0001 |
| 400 | 0.9981 | -0.0019 | 0.0000 | 0.9965 | -0.0035 | 0.0000 |
| 500 | 0.9986 | -0.0014 | 0.0000 | 0.9973 | -0.0027 | 0.0000 |
| 600 | 0.9989 | -0.0011 | 0.0000 | 0.9979 | -0.0021 | 0.0000 |
| 700 | 0.9992 | -0.0008 | 0.0000 | 0.9984 | -0.0016 | 0.0000 |
| 750 | 0.9991 | -0.0009 | 0.0000 | 0.9983 | -0.0017 | 0.0000 |

Table 2. Estimation results based on 1000 simulated series for various sample sizes of weighted LS method and LAD regression.

The estimated $\beta_1 (=1)$ using the two estimation methods are plotted in figure 2 for the different sample sizes. It can be seen that the weighted estimate performs better with respect to bias and the two estimators converges towards each other with an increase in the length of the series T.

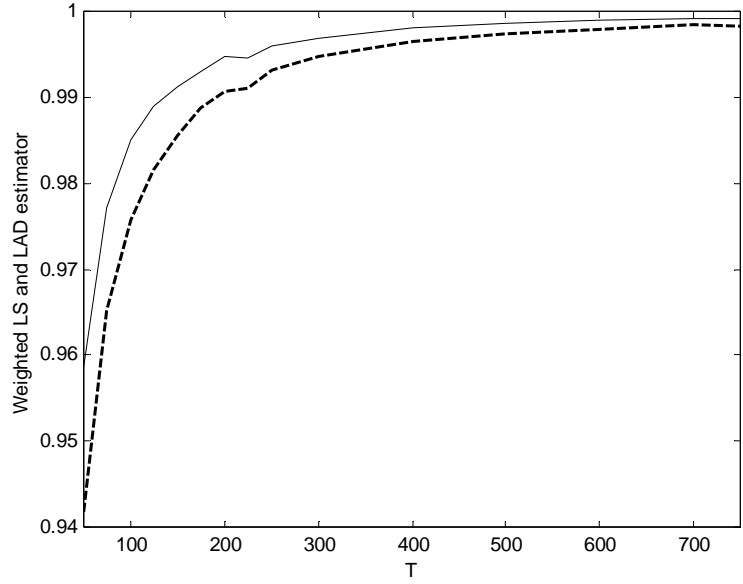

Figure 2. Plot of average estimated parameter using two estimation methods. Solid line denotes the mean weighted LS estimator, based on 1000 estimated parameters.

In table 3 results are given for estimating $\beta_1$ for the model $y_t = \beta_1 y_{t-1} + u_t$, series of length T=50, and 1000 series simulated each time. A series of length T=50 can be considered a small sample size in terms of time series. The weighted regression method outperforms LAD with respect to bias and MSE for unit root series or for $\beta_1$ close to one, and if the error term is heavy tailed distributed.

| $\beta_1$ | $\alpha$ | $\hat{\beta}_{1w}$ | $\hat{\beta}_{1L}$ |
|---|---|---|---|
| 1 | 2 | -0.0146 (0.0043) | -0.0322 (0.0047) |
| 1 | 1.9 | -.0099 (0.0034) | -0.0284 (0.0035) |
| 1 | 1.5 | -0.0024 (0.0011) | -0.0144 (0.0014) |
| 1 | 1.3 | -0.0025 | -0.0114 |

|       |     | (0.0006) | (0.0010)  |
|-------|-----|----------|-----------|
| 1     | 1.1 | -0.0008  | -0.0072   |
|       |     | (0.0002) | (0.0006)  |
| 0.975 | 2   | -0.0081  | -0.0318)  |
|       |     | (0.0053) | (0.0049)  |
| 0.975 | 1.9 | -0.0106  | -0.0302   |
|       |     | (0.0047) | (0.0045)  |
| 0.975 | 1.5 | -0.0050  | -0.0194   |
|       |     | (0.0019) | (0.0022)  |
| 0.975 | 1.3 | -0.0016  | -0.0118   |
|       |     | (0.0009) | (0.0011)  |
| 0.975 | 1.1 | 0.0006   | -0.0066   |
|       |     | (0.0004) | (0.0006)  |
| 0.5   | 2   | 0.0173   | -0.0196   |
|       |     | (0.0249) | (0.0147)  |
| 0.5   | 1.9 | 0.0194   | -0.0156   |
|       |     | (0.0244) | (0.0147)  |
| 0.5   | 1.5 | 0.0081   | -0.0140   |
|       |     | (0.0111) | (0.0079)  |
| 0.5   | 1.3 | 0.0088   | -0.0069   |
|       |     | (0.0069) | (0.0053)  |
| 0.5   | 1.1 | 0.0018   | -0.0058   |
|       |     | (0.0031) | (0.0033)  |

Table 3. Bias and MSE of model parameter using weighted regression and LAD regression with a stable distributed error term with different values of the index.

### 3.2 Unit root bootstrap test

In this section the testing of the unit root using bootstrap methods as given by Moreno and Romo (2012) will be investigated. The model that will be considered is $y_t = \beta_1 y_{t-1} + u_t$, $u_t$, stable distributed with index $0 < \alpha \leq 2$. The procedure has the following three steps:

a. Estimate $\beta_1$ from the original series $\{x_t\}, t = 1,...,T$, as $\hat{\beta}_1 = \sum_{t=2}^{T} x_t x_{t-1} / x_{t-1}^2$, using $\hat{\beta}_1$ calculate the residuals $\{\varepsilon_t\}$ and weights for the weighted estimate $\hat{\beta}_{1w} = \sum_{t=2}^{T} w_t x_t x_{t-1} / w_t x_{t-1}^2$. The weights are of the form

$$w_j = \exp(-|\hat{\varepsilon}_j - \hat{\varepsilon}_m|) / \sum_{j=1}^{T} \exp(-|\hat{\varepsilon}_j - \hat{\varepsilon}_m|), \; w_j, \hat{\varepsilon}_j, j = 1,...,T,$$

$\hat{\varepsilon}_m$ the median of the least squares residuals.

Calculate the residuals $\hat{\varepsilon}_{tw} = X_t - \hat{\beta}_{1w} X_{t-1}$, t=1,…,T. The test statistic to test the unit root hypothesis is $\max(|X_1|,...,|X_m|)\sqrt{T}(\hat{\beta}_{1w} - 1)$. making use of the result of Davis and Wu (1997), where it was shown that the random normalizing constant $\max(|X_1|,...,|X_m|)\sqrt{T}$ can be used in the original sample and in bootstrap samples.

b. Random bootstrap samples of size $m = T/\log(\log(T))$ is taken 4000 times, using the empirical distribution function of the $\{\hat{\varepsilon}_{tw}\}$. Generate series of length m under the null hypothesis, $X_t^* = X_{t-1}^* + \hat{\varepsilon}_t^*$, $\{\hat{\varepsilon}_t^*\}, t = 1,...,m$, the bootstrap sample. Calculate $\hat{\beta}_{1w}^*$ from this series and the test statistic $\max(|X_1^*|,...,|X_m^*|)\sqrt{m}(\hat{\beta}_{1w}^* - 1)$.

c. The test statistic obtained from the original series is the compared to the percentiles of the test statistics under the null hypothesis.

In the following table the results of a simulation to check the power of using the Moreno and Romo (2012) bootstrap procedure to test the unit root hypothesis is checked. The number of series simulated were 1000 each time, and to test the hypothesis 1000 bootstraps to estimate the critical value for testing the hypothesis: $H_0: \beta = 1.0$ with alternative $H_0: \beta < 1.0$.

|  | $\beta$ | | | |
|---|---|---|---|---|
| **n=50** | 0.8 | 0.9 | 0.95 | 1.00 |
| $\alpha = 2.0$ | 0.732 | 0.411 | .240 | .100 |
| $\alpha = 1.9$ | 0.784 | 0.486 | .256 | 0.084 |
| $\alpha = 1.5$ | 0.966 | .749 | .476 | 0.084 |
| $\alpha = 1.1$ | 1.000 | .961 | .800 | 0.087 |
| **n=250** | 0.8 | 0.9 | 0.95 | 1.00 |
| $\alpha = 2.0$ | 1.000 | 0.990 | 0.826 | 0.080 |
| $\alpha = 1.9$ | 1.000 | 0.997 | 0.874 | 0.094 |
| $\alpha = 1.5$ | 1.000 | 1.000 | .994 | 0.077 |
| $\alpha = 1.1$ | 1.000 | 1.000 | 1.00 | 0.120 |

Table 4. Power estimation when testing the unit root hypothesis using bootstrap methods

An indication of the performance of this estimator and the best performing one in in the results of Moreno and Romo (2012), for series with T=50 observations, $\alpha = 1.3$ is given in table 5. They compared hypothesis testing using LAD estimation, Tukey's bisquared function and other robust estimators. For this specific case Tukey's bisquared function performed best using the statistic suggested by Knight (1989), where a normalizing constant of the form $\left( \sum_{t=2}^{T} X_{t-1}^{T} \right)^{1/2}$, and also with bootstrap samples of size $m = T/\log(T)$ were sampled.

| $\beta$ | Tukeys's Bisquared Function | | Weighted Regression |
|---|---|---|---|
| | Knight | $m = T/\log(T)$ | |
| 0.8 | 97.35 | 94.50 | 99.20 |
| 0.9 | 82.35 | 77.05 | 90.40 |
| 0.95 | 55.45 | 50.15 | 63.00 |
| 1.0 | 6.65 | 5.35 | 8.40 |
| 1.01 | 27.20 | 23.65 | 5.50 |
| 1.05 | 86.90 | 87.25 | 90.00 |

Table 5. Comparison of the results of Moreno and Romo (2012) and using weighted estimation, bootstrap hypothesis testing.

## 4. Conclusions

The weighted estimation methods is very simple from the computational viewpoint and performs good when comparing with the LAD procedure which is more complicated from the computational viewpoint. This proposed procedure also yielded good results to test hypothesis using bootstrap methods.

For series with error terms which are not heavy-tailed distributed, very little is lost when using this procedure. Thus from a practical viewpoint, the weighted estimation might be very useful as an estimation procedure without having to use computational intensive methods.